\documentclass[preprint,showpacs,epsfig,eqsecnum,aps]{revtex4}
\usepackage{graphicx}
\usepackage{verbatim}
\begin{document}
\title{Companion problems in quasispin and isospin}

\author{L. Zamick and A. Escuderos}

\affiliation{Department of Physics and Astronomy, Rutgers University, 
Piscataway, New Jersey 08854}

\begin{abstract}
We note that the same mathematical results apply to problems involving 
quasispin and isospin, but the problems {\em per se} are different. In the 
quasispin case, one deals with a system of identical fermions (e.g. neutrons) 
and address the problem of how many seniority conserving interactions there 
are. In the isospin case, one deals with a system of both neutrons and protons 
and the problem in question is the number of neutron-proton pairs with a given 
total angular momentum. Other companion problems are also discussed.
\end{abstract}

\maketitle

\section{Introduction}
\label{sec:level1}

This work is based on the observation that the same mathematical results have 
been used to solve two different problems. One of the problems involves a 
system of identical fermions and the other, a system of mixed protons and
neutrons. In the first case, the quasispin formalism is used to address the
problem of the number of seniority conserving interactions there are, and in 
the second case, isospin considerations are used to simplify the expression for
the number of proton-neutron pairs of a given angular momentum. In both 
problems, it is found useful to obtain the eigenvalues and eigenfunctions of a 
unitary Racah coefficient. This can be done in a direct manner, but in both 
cases it is possible to obtain the eigenvalues without explicit 
diagonalization.

\section{First companion problem}
\label{sec:level2}

\subsection{Quasispin}
\label{sec:level21}

We here follow the works of Rosensteel and Rowe~\cite{rr01,rr03}. They have 
constructed `useful solvable and partially solvable shell model Hamiltonians' 
which conserve seniority. New numerical techniques were presented for computing
irreps of the USp$(2j+1)$ algebra. We will however focus on the part of 
their work where they find novel ways to solve problems which have been 
previously discussed and/or worked out by de Shalit and Talmi~\cite{st63},
Talmi~\cite{t93}, Lawson~\cite{l80}, French~\cite{f60}, French and 
MacFarlane~\cite{fm61}, Ginocchio and Haxton~\cite{gh93}---the 
number of seniority non-conserving interactions and constraints on 
seniority conserving interactions. This will be most relevant to the isospin 
part problem that we will later consider.

The quasispin formalism was introduced by Kerman~\cite{k61} and quickly 
developed by Kerman, Lawson and MacFarlane~\cite{klm6165}. For quasispin 
$S=\frac{1}{2}$ the operators are creation and destruction operators

\begin{equation}
\begin{array}{ll}
 (S,M_S) & \\
 (\frac{1}{2},\frac{1}{2}) & \hspace{1cm} a^+_{j m} \\
 (\frac{1}{2},-\frac{1}{2}) & \hspace{1cm} (-1)^{j-m} a_{j,-m} 
\end{array}
\end{equation}

We now suppress the single particle index $j$ in the following equations. For
$S=1$ we have

\begin{equation}
\begin{array}{ll}
(1,1) & \hspace{1cm} A_{LM}=\frac{1}{\sqrt{2}} [a^+a^+]_{LM} \\
(1,-1) & \hspace{1cm} B_{LM}=\frac{1}{\sqrt{2}} [a a]_{LM} \\
(1,0) & \hspace{1cm} C_{LM}=\frac{1}{2}\{[a^+ a]_{LM} + [a a^+]_{LM}\} = \\
 & \hspace{1.9cm} = [a^+ a]_{LM} - \sqrt{\frac{\Omega}{2}} \delta_{L,0}
\end{array}
\end{equation}

For the two particle interaction, we need quasispin operators of rank
0 and 2. The interaction can be written as

\begin{equation}
\hat{V}=-\frac{1}{4} \sum_{J \mbox{ \footnotesize even}} \sqrt{2J+1} 
V^J (A_J B_J)_0,
\end{equation}
where $V^J=\langle j^2,JM|V|j^2,JM \rangle$. For compactness the authors define
$Z_J=(A_J B_J)_0$.

The $S=0$ and $S=2$ quasispin operators with $M_S=0$ are

\begin{equation}
(111-1|S0) (A B)_0 + (11-11|S0) (B A)_0 + (1100|S0) (C C)_0.
\end{equation}
More specifically, Rosensteel and Rowe define them as

\begin{eqnarray}
X^0(J) & = & (A_J B_J)_0 - (C_J C_J)_0 + (B_J A_J)_0 \\
X^2(J) & = & (A_J B_J)_0 + 2(C_J C_J)_0 + (B_J A_J)_0 .
\end{eqnarray}

The six-$j$ symbol that we were referring to in the introduction enters when we
try to express the $X^S(J)$ in terms of the nucleon--nucleon interaction, or
more simply the $Z_J$. They find
\begin{eqnarray}
X^0(J) & = & -(M^\Omega -2I) Z_J + \cdots, \label{quasi-s0} \\
X^2(J) & = & 2(M^\Omega + I) Z_J + \cdots, \label{quasi-s2}
\end{eqnarray}
where $+\cdots$ refers to constants and terms linear in the number operator,
and $M^\Omega Z_J= \sum_\gamma Z_\gamma M^\Omega_{\gamma J}$, with

\begin{equation} 
M^\Omega_{\gamma J} = 2 \sqrt{(2J+1)(2\gamma +1)} \left\{
   \begin{array}{rrr}
   j & j & \gamma \\
   j & j & J
   \end{array} \right\}.
\end{equation}

They then state and prove `Proposition 1': the eigenvalues of $M^\Omega$ are
equal to $-1$ or $2$. They do so by noting that, if the eigenvalues were other
than those, then an eigenfunction of $M^\Omega$ would simultaneously have
quasispin $S=0$ and $S=2$, a contradiction. The projection operators for $S=0$ 
and $S=2$ are, then, respectively:

\begin{equation}
P_0=-\frac{1}{3}(M^\Omega -2I), \hspace{1.5cm} P_2=\frac{1}{3}(M^\Omega +I).
\end{equation}

They denote by $p_1$ the number of linearly independant rotationally invariant
quasispin scalars and by $p_2$, the quasispin rank 2 tensors. However, one of 
the $S=2$ operators is the pairing interaction $X^2_0(0)$ and this does not mix
states of different seniority. Hence, the number of seniority mixing 
interactions is $p_2-1$. They find that 
\begin{equation}
p_2=\mbox{\rm tr} P_2=\frac{1}{3}\left( \frac{2j+1}{2} + 
2 \sum_{\mbox{\footnotesize even }J} (2J+1)\left\{
   \begin{array}{ccc}
   j & j & J \\
   j & j & J
   \end{array} \right\} \right) = 
\left[\frac{2j+3}{6}\right], \label{p2}
\end{equation}
where the square brackets mean the integer part of what is inside. We will
discuss this more in the next section. We can see that eq.~(\ref{p2})
is the same result as Haxton and Ginocchio~\cite{gh93}, but using a 
novel technique. Rosensteel and Rowe~\cite{rr03} further obtain the condition 
that an interaction $\hat{V}$ should satisfy to conserve seniority:

\begin{equation}
(M^\Omega + I) V = \lambda (M^\Omega + I) Z_0.
\end{equation}
This yields the same conditions that are present in de Shalit and Talmi~\cite
{st63} and Talmi~\cite{t93}. We should emphasize that Rosensteel and 
Rowe~\cite{rr01,rr03} have other results which are new but that we are not
focusing on here.

\subsection{Isospin}
\label{sec:level22}

Some of the topics in this section have been discussed in part in 
preprints~\cite{zmsre-pp,zml-pp} and in conference 
proceedings~\cite{zelmmrs04}.

In the single $j$-shell model, the nucleus $^{44}$Ti consists of two valence 
protons and two valence neutrons. We can make an association of QUASISPIN and 
ISOSPIN by relating the ($S=\frac{1}{2}, M_S=\frac{1}{2}$) creation operator 
with a proton and the ($S=\frac{1}{2}, M_S=-\frac{1}{2}$) destruction operator 
with a neutron.

The wave function of a state $\alpha$ of total angular momentum $I$ can be
written as
\begin{equation}
\Psi^{\alpha I}=\sum_{J_P J_N} D^{\alpha I}(J_P J_N) [(j^2_\pi)J_P (j^2_\nu)
J_N]^I.
\end{equation}
Now we focus on $I=0$ states, for which $J_P=J_N\equiv J$ and

\begin{equation}
\Psi^{\alpha I=0}=\sum_{\mbox{\rm even $J$}} D^{\alpha I=0}(J J) [(j^2_\pi)J 
(j^2_\nu) J]^0.
\end{equation}
We already see a resemblance of the wave function for this problem with the
interaction for the quasispin problem. It can be shown that there are four
angular momentum $I=0$ states, three of which have isospin $T=0$ and one has
isospin $T=2$.

Now the $T=2$ state must be orthogonal to the $T=0$ states (for $I=0$):

\begin{equation}
\sum_J D^{T=2} (JJ) D^{\alpha T=0} (JJ)=0;
\end{equation}
we also have the normalization condition:

\begin{equation}
\sum_J D^{T=2} (JJ) D^{T=2} (JJ)=1.
\end{equation}
Since the $T=2$ state is the double analog of a state for a system of identical
nucleons (calcium isotopes), the unique $I=0, T=2$ wave function is known and
leads to the result

\begin{equation}
D^{I=0, T=2} (JJ)=(j^2 J j^2 J|\} j^4 0),
\end{equation}
where we have a two particle coefficient of fractional parentage (cfp) on the 
right hand side.

A useful identity can also relate the above to a one particle cfp:

\begin{equation}
D^{I=0, T=2} (JJ)=(j^2Jj|\}j^3j).
\end{equation}
However, a recursion formula, found on page 528 of de~Shalit and 
Talmi~\cite{st63}, namely, 
\begin{eqnarray}
\lefteqn{n(j^{n-1}(\alpha_0 J_0)j|\} j^n[\alpha_0 J_0] J)
(j^{n-1}(\alpha_1 J_1) j|\} j^n[\alpha_0 J_0] J) = }  \nonumber \\
 & = & \delta_{\alpha_1,\alpha_0} \delta_{J_1,J_0} + (n-1)
\sum_{\alpha_2 J_2} (-1)^{J_0+J_1} \sqrt{(2J_0+1)(2J_1+1)} \times \nonumber \\
 & & \times \left\{
   \begin{array}{ccc}
   J_2 & j & J_1 \\
   J & j & J_0
   \end{array} \right\}
(j^{n-2}(\alpha_2 J_2) j|\} j^{n-1} \alpha_0 J_0)
(j^{n-2} (\alpha_2 J_2) j|\} j^{n-1} \alpha_1 J_1), \label{recursion}
\end{eqnarray}
with $n$ set equal to 3, leads to the following result, valid
for $j$ values of $3/2,5/2$ and $7/2$:
\begin{equation}
\sqrt{(2J+1)(2J'+1)} \left\{
   \begin{array}{lll}
   j & j & J' \\
   j & j & J
   \end{array} \right\} =
-\frac{\delta_{JJ'}}{2} + \frac{3}{2} (j^2(J)j|\}j^3 j v=1)
(j^2(J')j|\}j^3 j v=1). \label{rec-cfp}
\end{equation}
This leads to the following:

\begin{eqnarray}
2\sum_{J_P} \sqrt{(2J_P +1)(2J_{12} +1)} \left\{
   \begin{array}{lll}
   j & j & J_P \\
   j & j & J_{12}
   \end{array} \right\} D(J_P J_P) & = & 
-D(J_{12} J_{12}) \hspace{.6cm} \mbox{\rm for $T=0$} \label{iso-t0} \\
 & = & 2 D(J_{12} J_{12}) \hspace{.6cm} \mbox{\rm for $T=2$}. \label{iso-t2}
\end{eqnarray}
But this is an eigenvalue problem for the unitary $6j$-symbol and the 
eigenvalues are now shown to be $-1$ and $2$ without explicit diagonalization.
These are the same eigenvalues for the same operator that Rosensteel and Rowe
state in their Proposition 1~\cite{rr01,rr03}. The equations~(\ref{iso-t0}) and
(\ref{iso-t2}) have the same physical structure as~(\ref{quasi-s0}) and 
(\ref{quasi-s2}).

In the isospin case, the vectors are the wave function components $D(JJ)$; in 
the quasispin
case, the vectors are the $Z(J)$'s, i.e., $\left[ [a^+a^+]^J [aa]^J\right]^0$.
In the isospin problem, the practical application of this is to obtain the 
number of neutron--proton pairs in $^{44}$Ti. The general expression is 
complicated, but for even $J_{12}$ of a pair, the result using the $6j$
eigenvalue equation for $^{44}$Ti is
\begin{equation}
\mbox{\rm \# of nn pairs = \# of np pairs = \# of pp pairs }=
|D(J_{12} J_{12})|^2,
\end{equation}
with $J_{12}=0,2,4$ and 6. Simple expressions for the number of $J_{12}=0$ 
pairs for $^{46}$Ti and $^{48}$Ti have also been obtained~\cite{zmsre-pp}.

We can use eq.~(\ref{rec-cfp}) to cast some limited insight into the result of 
Rosensteel and Rowe~[eq.~(\ref{p2})]. This result involve
\begin{equation}
\mbox{SUM}6j=\sum_{\mbox{\footnotesize even }J} (2J+1) \left\{
   \begin{array}{ccc}
   j & j & J \\
   j & j & J
   \end{array} \right\}.
\end{equation}
For $j\leq 7/2$ we can use eq.~(\ref{rec-cfp}) to evaluate this. We have the
following result for the cfp's:
\begin{equation}
\sum_J |(j^2(J)j|\}j^3 j)|^2 = 1.
\end{equation}
This condition comes from the fact that the wave function for three identical
particles is normalized to unity. Hence, for $j=3/2,5/2$ and7/2, we have the
following result:

\begin{equation}
\mbox{SUM}6j=-\frac{2j+1}{4} + \frac{3}{2}, \label{sum6j}
\end{equation}
i.e., $0.5,0$ and $-0.5$ for the three $j$ values above. This formula does not
work for $j=1/2$ because we cannot have 3 neutrons in the $s_{1/2}$ shell; we
can, however, use the same recursion formula~(\ref{recursion}) with $n=2$ in
this case to find that SUM$6j=-0.5$ (as one can just look it up). 
Eq.~(\ref{sum6j})
does not work for $j=9/2$ or higher because in that case there is more than
one state of a given angular momentum; e.g., for $J=j=9/2$ there are two 
states, one with seniority 1 and the other with seniority 3.

A more general result has been obtained by Zhao, Arima, Ginocchio and 
Yoshinaga~\cite{zagy03}. The pattern ($-0.5, 0.5, 0$) for $j=s_{1/2},
p_{3/2}$ and $d_{5/2}$ respectively repeats itself, i.e., the same trio
of results holds for ($f_{7/2}, g_{9/2}, h_{11/2}$), etc. This was shown by 
the above authors~\cite{zagy03} simply by equating the left-hand side of 
eq.~(\ref{p2}) with
the right-hand side, and relying on the proof in ref.~\cite{gh93} that
$\left[\frac{2j+1}{6}\right]$ is indeed the number of states of different 
seniority. For completeness, we give also two other references by this group 
(\cite{za03}, \cite{za04}).

(Whereas the sum  in SUM$6j$ is over even angular momenta, it was pointed
out to us by I.~Talmi that the sum over all angular momenta is much easier 
to obtain. Indeed the result is explicitly given in a work by 
J.~Schwinger ``On Angular Momentum'', eqs.~(4.27) and (4.28). The sum over all 
J is zero for half-integer $j$ and one for integer $j$~\cite{s65}).

We can use {\it isospin} arguments to derive the left hand side 
Rosensteel--Rowe relation~[eq.~(\ref{p2})]. We consider a system of one proton 
and two neutrons in a single $j$-shell.

We use isospin variables $p$ and $n$ for the proton and neutron. The basis
states are, then,

\begin{equation}
\psi^I [J_0] = \frac{1}{\sqrt{3}} (1-P_{12} -P_{13}) \left[ j(1) 
[j(2)j(3)]^{J_0} \right]^I p(1) n(2) n(3).
\end{equation}
We introduce a simplified hamiltonian $V=\sum_{i<j} (a+b t(i)\cdot t(j))$, 
where $a$ and $b$ are constants.

We now evaluate the trace
\begin{equation}
\mbox{tr} [J_0] = \sum_{J_0 \mbox{ \scriptsize  even}} \langle \psi^I 
[J_0] V \psi^I [J_0] \rangle,
\end{equation}
and we find
\begin{eqnarray}
\mbox{tr} [J_0] & = & \sum_{J_0 \mbox{ \scriptsize  even}} \left[ 
\left(3a-\frac{b}{4}\right) + b (2J_0 +1) \left\{
   \begin{array}{ccc}
   j & j & J_0 \\
   j & j & J_0
   \end{array} \right\}
\right] = \nonumber \\
 & = & \left(3a-\frac{b}{4}\right) \frac{2j+1}{2} + b \sum_{J_0 \mbox{ 
\scriptsize  even}} (2J_0 +1) \left\{
   \begin{array}{ccc}
   j & j & J_0 \\
   j & j & J_0
   \end{array} \right\} . 
\end{eqnarray}

The expectation value of $V$ for an $A$ body system with total isospin $T$ is:

\begin{equation}
\langle V \rangle = \frac{A(A-1)}{2}a + \frac{b}{2} T(T+1) - \frac{3}{8}bA,
\end{equation}
which, for $A=3$, becomes
\begin{equation}
\langle V \rangle = 3a + \frac{b}{2} T(T+1) - \frac{9}{8} b.
\end{equation}
We choose $a$ and $b$ so that $\langle V \rangle =0$ for $T=1/2$ and
$\langle V \rangle =1$ for $T=3/2$. We find $a=1/6$, $b=2/3$. With this choice,
$\mbox{tr}[J_0]$ becomes the number of $T=3/2$ states of angular momentum $I$,
which is also the same as the number of states of angular momentum $I$ for a
system of 3 identical particles.

We then obtain $\mbox{tr} [J_0] = p_2$, as in eq.~(\ref{p2}), namely
\begin{equation}
\frac{1}{3}\left( \frac{2j+1}{2} + 
2 \sum_{J_0\mbox{ \scriptsize even}} (2J_0+1)\left\{
   \begin{array}{ccc}
   j & j & J_0 \\
   j & j & J_0
   \end{array} \right\} \right) . \label{p2-2}
\end{equation}
The above result supports the theme of this work that there is an 
interrelationship between quasispin and isospin.

\section{Second companion problem}
\label{sec:level3}

In the realm of identical fermions, e.g. neutrons, the Pauli Principle imposes
a severe restriction on the number of states that are allowed. For example, in
the calcium isotopes, if we limit ourselves to one single $j$-shell 
$j=f_{7/2}$, then the allowed states for $^{43}$Ca are $I=3/2, 5/2, 7/2, 9/2,
11/2$ and $15/2$, all occurring only once; while for $^{44}$Ca the allowed 
angular momentum--seniority combinations are $I=0, v=0$, $I=2,4,6$ all with
$v=2$, and $I=2,4,5$ and 8 with seniority $v=4$. All states of given ($I,v$)
occur only once.

As discussed in de~Shalit and Talmi~\cite{st63} and Talmi~\cite{t93}, if one is
foolish enough to try to calculate a coefficient of fractional parentage for a
non-existent state, one gets zero. But this can produce useful results. For
example, in $^{43}$Ca there is no $I=13/2$ state of the $f_{7/2}^3$ 
configuration. This leads to the result for a certain $6j$-symbol:

\begin{equation}
\left\{
  \begin{array}{ccc}
  \displaystyle \frac{7}{2} & \displaystyle \frac{7}{2} & 4 \vspace{0.25cm} \\
  \displaystyle \frac{13}{2} & \displaystyle \frac{7}{2} & 6 
  \end{array}
\right\} =0.
\end{equation}
This result can be easily generalized to

\begin{equation}
\left\{
  \begin{array}{ccc}
  j & \hspace{0.3cm} j \hspace{0.3cm} & (2j-3) \\
  (3j-4) & \hspace{0.3cm} j \hspace{0.3cm} & (2j-1) 
  \end{array}
\right\} =0.
\end{equation}

The companion problem deals with the isospin $T=\frac{1}{2}$ states in 
$^{43}$Sc (or the mirror $^{43}$Ti). This is in contrast to the original 
problem which deals with $T=\frac{3}{2}$ states in $^{43}$Ca. For $^{43}$Sc one
requires knowledge of both the proton--neutron interaction and the 
neutron--neutron interaction. In terms of isospin, the $nn$ system must have
isospin 1, but we can have both $T=0$ and $T=1$ $np$ states. Indeed, in the
single $j$-shell the $np$ states with even total angular momenta $J$ have
isospin $T=1$, whereas those with odd $J$ value have $T=0$.

Robinson and Zamick~\cite{rz01-63,rz01-64,rz02} posed the question of what 
happens if the two-body $T=0$ matrix elements are set equal to zero (or, what 
amounts to the same thing, a constant). The motivation is contained in the 
references and will not be repeated here. It was found that the result of this 
was twofold.

In $^{43}$Sc the basis states can be classified by ($J_N,j$), the angular 
momentum of the two neutrons and that of the single proton. The wave function
can be written $\Psi^{\alpha I} =\sum_{J_N\mbox{\rm\footnotesize  even}}
C^{\alpha I}(J_N,j) [(j^2)^{J_N} j]^I$. For a certain set of states, a 
dynamical symmetry was found. These $T=\frac{1}{2}$ states had angular momenta
$I=1/2,13/2,17/2$ and 19/2. For these states, when the two-body matrix elements
were set to zero, the wave functions had quantum numbers ($J_N,j$), i.e., one
13/2 state was (4,7/2) and the other (6,7/2). Furthermore, there were
degeneracies. The $I=1/2$ and 13/2 states were degenerate as were the 17/2 and
19/2 states.

How does this fact relate to the original $^{43}$Ca problem? When we ask why
the matrix element $\langle [4,7/2]^{13/2} V [6,7/2]^{13/2} \rangle$ vanishes,
we find that the expression involves the $6j$-symbol $\left\{ 
\begin{array}{ccc} 7/2 & 7/2 & 4 \\ 13/2 & 7/2 & 6 \end{array}\right\}$, which
was shown to vanish because there was no 13/2 state of the $f_{7/2}^3$ 
configuration in $^{43}$Ca. The vanishing of the 13/2 cfp also leads to a 
diagonal condition which leads to the degeneracy of the states $[4,7/2]^{13/2}$
and $[4,7/2]^{1/2}$, as well as $[6,7/2]^{17/2}$ and $[6,7/2]^{19/2}$.

Note that we have a partial dynamical symmetry. States of the other angular 
momenta, i.e. $I=3/2,5/2,7/2,9/2,11/2$ and 15/2 do not behave in this way. We 
now see what is happening---this partial dynamical symmetry for $T=1/2$ states
in $^{43}$Sc applies only to angular momenta which do not occur for the 
$f_{7/2}^3$ configuration of $^{43}$Ca.

We can see this more clearly by comparing the $I=13/2$ and $I=15/2$ states in
$^{43}$Sc. For both cases the basis states are $[4,7/2]^I$ and $[6,7/2]^I$.
For $I=13/2$, both states have $T=1/2$. However, for $I=15/2$, one state has 
$T=1/2$ and one has $T=3/2$. The latter state is an analog state of the 
unique $I=15/2$ state in $^{43}$Ca. For $I=15/2$ the two states can be written
as

\begin{eqnarray}
\Psi_1 & = & a [4,7/2] + b [6,7/2] \\
\Psi_2 & = & -b [4,7/2] + a [6,7/2].
\end{eqnarray}
Let the first state be the $T=3/2$ state. Because it is the analog of a state 
in $^{43}$Ca, we can easily show that $a$ and $b$ are coefficients of
fractional parentage:

\begin{eqnarray}
a & = & (j^2 4 j |\} j^3 15/2) = \sqrt{\frac{5}{22}}, \\
b & = & (j^2 6 j |\} j^3 15/2) = \sqrt{\frac{17}{22}}.
\end{eqnarray}
This is independant of what (isospin conserving) interaction is chosen. The
second state, with isospin $T=1/2$, must be orthogonal to the first $T=3/2$
state, so its components also involve the same cfp's. The point is that here
also the interaction cannot change the wave function, and so we do not get a
partial dynamical symmetry.

To summarize this section, the vanishing of a $6j$-symbol enters into two
companion problems---why certain angular momenta cannot occur for an 
($f_{7/2}^3$) $T=3/2$ configuration and why, for these same angular momenta,
the $T=1/2$ states can be classified by the quantum numbers ($J_N,j$), and
further why states with the same ($J_N,j$) but different angular momenta are
degenerate.

As a brief summary of both sections, we have shown that the same mathematical
relations can be used to solve problems involving systems of neutrons and
protons, and systems of identical particles. With the exception of 
the result of eq.~(\ref{p2-2}), the problems are quite different. Our 
distinctive contributions have been for the systems of mixed neutrons and
protons involving isospin and we have established connections with the work
on identical particles of Haxton, Ginocchio, Rosensteel, Rowe, Zhao, Arima,
and Yoshinaga, as well as de~Shalit, Talmi and Racah. Perhaps in the near 
future other examples will emerge.

We thank David Rowe for pointing out his very interesting work to us, and Igal
Talmi for very useful communications. One of us (A.E.) is supported by a grant 
financed by the Secretar\'\i a de Estado de Educaci\'on y Universidades (Spain)
and cofinanced by the European Social Fund. We also acknowledge support of the
U.S. Dept. of Energy under contract No. DE-FG0104ER04-02.



\vfill\eject



\begin{thebibliography}{00}
\bibitem{rr01} D.J. Rowe and G. Rosensteel, Phys. Rev. Letters {\bf 87} (2001)
172501.

\bibitem{rr03} G. Rosensteel and D.J. Rowe, Phys. Rev. {\bf C67} (2003) 014303.

\bibitem{st63} A. de Shalit and I. Talmi, {\it Nuclear Shell Theory}, Academic
Press, New York (1963).

\bibitem{t93} I. Talmi, {\it Simple Models of Complex Nuclei}, Harwood 
Academic, Switzerland (1993).

\bibitem{l80} R.D. Lawson, {\it Theory of the Nuclear Shell Model}, Clarendon 
Press, Oxford (1980).

\bibitem{f60} J.B. French, Nucl. Phys. {\bf 15} (1960) 393.

\bibitem{fm61} J.B. French and M.H. Macfarlane, Nucl. Phys. {\bf 26} (1961) 
168.

\bibitem{gh93} J.N. Ginocchio and W.C. Haxton, {\it Symmetries in Science VI},
ed. by B.~Gruber and M.~Ramek, Plenum, New York (1993).

\bibitem{k61} A.K. Kerman, Ann. Phys. (N.Y.) {\bf 12} (1961) 300.

\bibitem{klm6165} A.K. Kerman, R.D. Lawson and M.H.~MacFarlane, Phys. Rev. 
{\bf 124} (1961) 162; Nucl. Phys. {\bf 66} (1965) 80.

\bibitem{zmsre-pp} L.~Zamick, E.~Moya de~Guerra, P.~Sarriguren, A.A.~Raduta
and A.~Escuderos, LANL preprint, nucl-th/0312077.

\bibitem{zml-pp} L.~Zamick, A.Z.~Mekjian and S.J.~Lee, LANL preprint,
nucl-th/0402089.

\bibitem{zelmmrs04} L.~Zamick, A.~Escuderos, S.J.~Lee, A.~Mekjian, E.~Moya
de~Guerra, A.A.~Raduta and P.~Sarri\-guren, Expressions for the Number of Pairs
of a Given Angular Momentum in the Single $j$ Shell Model, {\it Proceedings 
of the 8th International Spring Seminar on Nuclear Physics, Key Topics in 
Nuclear Structure}, Paestum (Italy), May 23--27, 2004.

\bibitem{rz01-63} S.J.Q.~Robinson and L.~Zamick, Phys. Rev. {\bf C63} (2001) 
064316

\bibitem{rz01-64} S.J.Q. Robinson and L. Zamick, Phys. Rev. {\bf C64} (2001)
057302

\bibitem{rz02} S.J.Q. Robinson and L.~Zamick, Fermionic Symmetries in Nuclei,
{\it Proceedings of the 7th International Spring Seminar on Nuclear Physics.
Challenges of Nuclear Structure} edited by A.~Covello, World Scientific (2002) 
223--230.

\bibitem{zagy03} Y.M.~Zhao, A.~Arima, J.N.~Ginocchio and N.~Yoshinaga, Phys.
Rev. {\bf C68} (2003) 044320.

\bibitem{za03} Y.M.~Zhao and A.~Arima, Phys. Rev. {\bf C68} (2003) 044310.

\bibitem{za04} Y.M.~Zhao and A.~Arima, Phys. Rev. {\bf C70} (2004) 034306.

\bibitem{s65}  J.~Schwinger, On Angular Momentum, published in {\it Quantum 
Theory of Angular Momentum}, ed. by L.C.~Biedenharn and H.~Van~Dam, Academic 
Press, New York (1965) 229--279.


\end{thebibliography}
\end{document}